# Effects of the interplay between the orbital and spin currents in electroexcitation of light nuclei

E.R. Arakelyan, N.G.Goncharova*

*Department of Physics, Lomonosov Moscow State University, Moscow 119992, Russia*

**Abstract.** The interplay between the orbital and spin currents in $(e,e')$ excitations of light nuclei is discussed. The microscopic analysis of $E1$ and $M2$ resonances was performed within the particle-core coupling approximation (PCC) of the shell model. The comparison of the theoretical results with the available experimental data has shown that the studies of relative contributions of the orbital and different spin components of the nuclear current into the excitation of $EJ$ and $MJ$ resonances could be useful for the identification of their configurational structure.

## 1. Introduction

The multipole giant resonances (MGR) are observed in a variety of nuclear reactions when an incident particle interacts with the nuclear charge or nuclear convection currents. The most detailed investigation of MGR has been performed in the photonuclear reactions and the reactions of inelastic electron scattering $(e,e')$ from nuclei because the analysis of these reactions allows us to separate the effects of the interaction of a real or virtual photon with the nuclear charge from those due to its interaction with the nucleon currents within the nucleus. The nucleon currents involved in the process of nuclear electroexcitation in $(e,e')$ reactions are caused by both the motion of the charges within the nucleus (orbital currents) and the reorientation of the magnetic moments of the nucleons (spin currents). Meson exchange currents play a relatively small role in nuclear electroexcitations at the momentum transfer $q$ below ~ 500 MeV/c.

Studies of MGR in nuclei via inelastic electron scattering make it possible, by varying independently the momentum transfer $q$ and the scattering angle $\theta$, to observe both spin- and orbital-current contributions to nuclear excitations. The nuclear structure affects the $(e,e')$ cross section through the longitudinal and transverse form factors $F_L^2$ and $F_T^2$ [1]:

$$\frac{d^2\sigma(e,e')}{d\Omega d\omega} = \frac{4\pi\sigma_M}{\eta_R}\left\{\left(\frac{q_\mu^4}{q^4}\right)F_L^2(q,\omega) + \left(\frac{q_\mu^2}{2q^2} + tg^2\frac{\theta}{2}\right)F_T^2(q,\omega)\right\} \quad , \qquad (1)$$

* *e-mail*: n.g.goncharova@physics.msu.ru

$$F_L^2(q) = \sum_0^{J=J_{max}} F_{CJ}^2(q) = (2J_i+1)^{-1} \sum_J \left| \left\langle J_f \left\| \hat{M}_J^{coul} \right\| J_i \right\rangle \right|^2 \quad , \tag{2}$$

$$F_T^2(q) = \sum_1^{J_{max}} (F_{EJ}^2 + F_{MJ}^2) = (2J_i+1)^{-1} \sum_J \left\{ \left| \left\langle J_f \left\| \hat{T}_J^{el} \right\| J_i \right\rangle \right|^2 + \left| \left\langle J_f \left\| \hat{T}_J^{mag} \right\| J_i \right\rangle \right|^2 \right\} \quad . \tag{3}$$

$J_i$ and $J_f$ in Eq. (2),(3) are the spins of the nucleus $A$ in the initial and final states. The Coulomb operator $\hat{M}_J^{coul}$ in Eq.(2) represents the nuclear charge density, the electric and magnetic operators $\hat{T}_J^{el}$ and $\hat{T}_J^{mag}$ in Eq. (3) contain contributions from the spin and orbital nuclear currents.

In a shell model picture, $\hat{T}_J^{el}$ and $\hat{T}_J^{mag}$ can be written as a sum of one-body operators where the operator $[Y_J \times \hat{\sigma}_i]$ is associated with the spin current, and the operators $[Y_{J-1} \times \hat{\nabla}_i]$ and $[Y_{J+1} \times \hat{\nabla}_i]$ reflect the orbital-current contributions to the transverse form factors of $EJ$ and $MJ$ excitations:

$$
\begin{aligned}
\hat{T}_{JM}^{el} = \frac{q}{2M} \sum_{i=1}^A \Bigg\{ & \hat{g}_i \, j_J(qr_i) [Y_J(\Omega_i) \times \hat{\sigma}_i]^{JM} \\
& + \frac{2\hat{e}_i}{q} \left( \sqrt{\frac{J+1}{2J+1}} \, j_{J-1}(qr_i) [Y_{J-1}(\Omega_i) \times \hat{\nabla}_i]^{JM} - \sqrt{\frac{J}{2J+1}} \, j_{J+1}(qr_i) [Y_{J+1}(\Omega_i) \times \hat{\nabla}_i]^{JM} \right) \Bigg\} \\
= & \; \hat{A}_J + \hat{B}_{J+1} + \hat{B}_{J-1} \quad ,
\end{aligned}
\tag{4}
$$

$$
\begin{aligned}
\hat{T}_{JM}^{mag} = \frac{iq}{2M} \sum_{i=1}^A \Bigg\{ & \left( \hat{g}_i \sqrt{\frac{J+1}{2J+1}} \, j_{J-1}(qr_i) [Y_{J-1}(\Omega_i) \times \hat{\sigma}_i]^{JM} - \sqrt{\frac{J}{2J+1}} \, j_{J+1}(qr_i) [Y_{J+1}(\Omega_i) \times \hat{\sigma}_i]^{JM} \right) \\
& - \frac{2\hat{e}_i}{q} \left( j_J(qr_i) [Y_J(\Omega_i) \times \hat{\nabla}_i]^{JM} \right) \Bigg\} = \hat{A}_{J-1} + \hat{A}_{J+1} + \hat{B}_J \quad .
\end{aligned}
\tag{5}
$$

In (4), (5) $\hat{A}_J \in j_J(qr)[Y_J \times \hat{\sigma}]^J$ and $\hat{B}_J \in j_J(qr)[Y_J \times \hat{\nabla}]^J$ .

The separation of the spin-angular variables from the radial ones (represented by the spherical Bessel functions $j_J(qr)$ ) is justified by the fact that the spin and angular dependences of the matrix elements of the single-particle operators are universal for the electromagnetic, strong and weak interactions, while the radial dependence of these matrix elements is determined by the specific dynamics of a particular reaction.

The operators $\hat{e}_i, \hat{g}_i$ work in the isospin space and therefore can be either isoscalar ($T=0$) or isovector ($T=1$). The former corresponds to the interaction of a real or virtual photon with nucleon convection currents, the latter (so-called "$g$-factor") corresponds to the magnetization currents. The isoscalar and isovector operators $\hat{e}_i$ have the same magnitude: $\hat{e}_{T=1} = \hat{e}_{T=0} = 0.5$, but the magnitude of the isovector operator $\hat{g}_{T=1}$ in the case of "free" nucleons is more than 5 times larger than that of the isoscalar operator: $\hat{g}_{T=1} = 2.35, \hat{g}_{T=0} = 0.44$. The effects of the interference between the spin- and orbital-current contributions into the ($e,e'$) reaction cross sections are most explicitly manifested in the isovector ($T=1$) nuclear excitations which will be discussed in detail below.

It should be noted that, according to the studies of relative contributions from the spin and orbital currents into the ($e,e'$) cross section, the isovector $g$-factor applied to the nuclear matter (as opposed to the $g$-factor for "free" nucleons $g_{free}$) has to be renormalized as follows: $g = X \times g_{free}$, where the renormalization factor $X$ is usually assumed to be 0.7. Only after this renormalization a good agreement between the theory and the experimental data can be achieved. To what extent the value of $X$ depends on the specific nucleus still remains an open question

The multipolarity $J$ of the excitation operators in ($e,e'$) reactions varies with growing momentum transfer $q$. As an example, Fig.1 shows the summed transversal form factors for the isovector $EJ$ and $MJ$ resonances in $1\hbar\omega$ excitations of the $sd$-shell nucleus $^{28}$Si [2]. The calculations were performed using the harmonic-oscillator wave functions (HOWF).

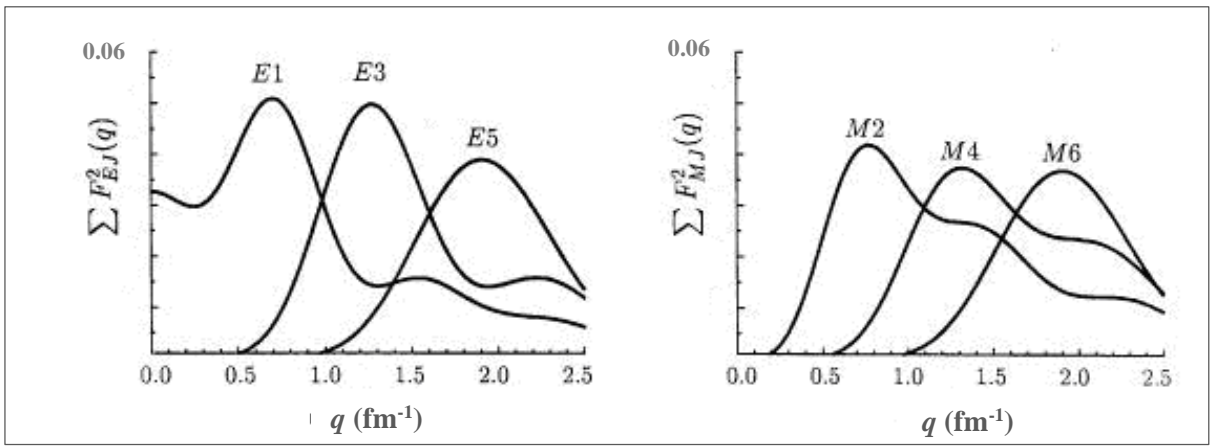

**Fig. 1.** Transverse form factors $EJ$ and $MJ$ for the isovector $1\hbar\omega$ transitions in $^{28}$Si [2].

The increase in the momentum transferred to the nucleus leads to a growing role of the spin currents in the excitation of MGR. The transverse $E1$ form factors at the "photopoint" $q = E_{excitation}$ include only the orbital-current contributions. The maximum spins in $1\hbar\omega$ nuclear excitations appear in the so-called "stretched states" which in the case of $sd$-shell nuclei at the momentum transfer $q \sim 2$ fm$^{-1}$ correspond to $M6$ resonances. The excitation of these $M6$ "stretched states" is solely due to the spin component of the nucleon current: $\hat{A}_5 \in j_5(qr)[Y_5 \times \hat{\sigma}]^6$.

## 2. Model input

GMR are the universal property of nuclear response to excitation. They represent the most striking feature of the reaction cross sections for the excitation energies $E < 40$ MeV and momentum transfers $q < 2.5$ fm$^{-1}$.

The interpretation of the GMR structure and its dependence on the individual properties of a specific nucleus is one of the main goals of nuclear theory. The theoretical description of the magnitudes, energy distributions and partial characteristics of GMR within the multi-particle shell model (MSM) has been developed for the past 50 years. The growing scope of the experimental information on the GMR structure has shown that the MSM calculations based on the particle-hole configurations are unable to correctly reproduce the complicated shapes of GMR. A usual way to overcome this problem is to expand the space of basic configurations by taking into account the interaction of the "doorway" states with more complex configurations, and, first of all, with collective phonons. Applications of this method to the GMR in the medium and heavy closed-shell nuclei were quite successful, but the interpretation of the GMR structure and decay properties in the open-shell nuclei, especially in the deformed ones, still represents a challenge to the theory. The excitation of nuclei also leads to dynamical deformations of nuclear systems that have near-spherical shapes in the ground state. How to take all these effects into account is still an open question. Moreover, the so-called "magic" nuclei cannot be considered as bona fide closed-shell systems due to the pairing forces in nuclei.

One of the possible ways to build a set of basic configurations that could be used as "doorway" states in the microscopic description of GMR in open-shell nuclei is taking into account the distribution of the "hole" configurations among the states of the residual ($A$-1) nuclei. This method was used in the particle-core coupling approach of the shell model (PCC SM) [3]. The deviation of a nucleus $A$ from the closed shells or subshells reveals itself in a wide range of energy distributions for the states of ($A$-$I$) nuclei.

The wave functions $\left| J_f T_f \right\rangle$ of the excited states for the nucleus $A$ in this approach are expanded into a set of low-lying states $(J'E'T')$ of the residual nuclei ($A$-1) coupled with a nucleon in a free orbit $(n'l'j')$:

$$\left| J_f T_f \right\rangle = \sum_{(J'),j'} \alpha_f^{J'j'} \left| (J'E'T')_{A-1} \times (n'l'j'):J_f T_f \right\rangle \quad . \tag{6}$$

The coefficients $\alpha$ are obtained by diagonalizing the PCC Hamiltonian on the PCC configuration basis (6). The "core" states $(J'E'T')_{A-1}$ include all the states of the nuclei ($A$-1) with a non-vanishing fractional-parentage relation to the ground state of the nucleus $A$:

$$\left| J_i T_i \right\rangle = \sum_{(J'),j} C_i^{(J'),j} \left| (J'E'T')_{A-1} \times (nlj):J_i T_i \right\rangle \quad , \tag{7}$$

where $C_i$ are fractional parentage coefficients.

The matrix elements of the operators in the space of configurations (2), (3) could be represented as sums of the matrix elements for the single-particle transitions multiplied by the spectroscopic amplitudes $Z$:

$$\left\langle J_f T_f M_T \left\| \hat{B}_{T M_T}^J \right\| J_i T_i M_T \right\rangle = \sum_{i, j_i, j_f} \left\langle j_f \left\| \hat{O}_{T M_T}^J \right\| j_i \right\rangle \sqrt{2 \cdot (2J_i + 1)} \cdot Z_{T M_T}^J (j_f j_i) \quad . \tag{8}$$

The information on the structure of the initial and final states of the nucleus $A$ in (8) is embedded in the spectroscopic amplitudes $Z$:

$$Z_{T M_T}^J (j_f j_i) = \sqrt{(2T+1)(2T_i+1)(2J_f+1)} \left\langle T_i M_T T 0 \middle| T_f M_T \right\rangle$$

$$\times \sum_{J_i T'} C_i^{JT,j_i} \; \alpha_f^{J'T',j_f} (-1)^{J'-J_i+j_f-J} \, W(J_i J_f j_i j_f; J J') (-1)^{T'-T_i+\frac{1}{2}-T} W\left(T_i T_f \frac{1}{2}\frac{1}{2}; TT'\right). \quad (9)$$

The matrix elements of the PCC Hamiltonian involve the excitation energies of the energy levels of final nuclei:

$$\hat{H}_{ij} = (E' + \varepsilon_j + E_c)\delta_{ij} + \hat{V}_{ij} \quad . \qquad (10)$$

The estimation of the magnitudes of the residual interaction matrix elements was also based on the probabilities of pick-up reactions according to the following scheme:

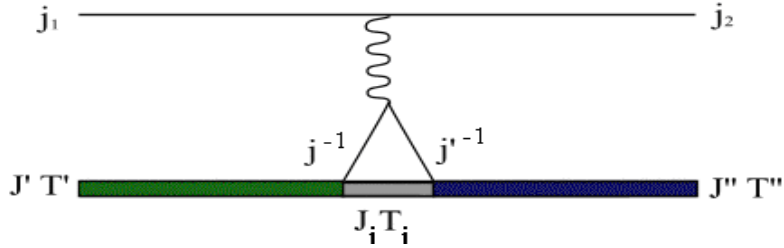

The PCC basis can easily be extended by including the states $(J'E'T')_{A-1}$ coupled with collective excitations of the target nucleus $A$. The PCC SM approach also allows us to obtain in a relatively simple way the decay characteristics of GMR.

The fractional parentage coefficients $C_i$ are the result of the expansion of the ground-state wave function for the nucleus $A$ into a set of wave functions for the final nucleus $(A-1)$ coupled with the wave function of a "free" nucleon. For $1p$-shell nuclei the ground-state wave functions [4] used in the calculations of the fractional parentage coefficients $C_i$ were obtained within SM with intermediate coupling.

For the nuclei with $A > 16$ the evaluation of $C_i$ was performed using the spectroscopy data on the direct nucleon pick-up reactions:

$$\tilde{N}_i = \sqrt{S_i \Big/ \sum S_i} \; ,$$

where $S_i$ is the spectroscopic factor for the reaction leading to the excitation of the $(J'E'T')$ level in the final nucleus $(A-1)$.

## 3. $E1$: Electric dipole excitations

Electric dipole resonance in $(e,e')$ reactions reveals itself in the forward electron scattering as longitudinal $C1$ form factor and in the backward scattering as transverse $E1$ form factor. The shapes of these form factors are quite similar in the vicinity of the photopoint ($q = \omega$), in accordance with the Siegert's theorem [1]. However, this similarity is vanishing with growing momentum transfer $q$. Form factors for all single-particle longitudinal $CJ$ transitions of different multipolarities $J$ have virtually the same $q$-dependences. At the same time, the behavior of the respective transverse $EJ$ transitions is more complicated as the interference between the orbital and spin nuclear currents participating in their formation may be either constructive or destructive.

The analysis of the isovector $1\hbar\omega$ $E1$ transitions $1l_{j=l+1/2} \rightarrow 1(l+1)_{j=l+3/2}$ has shown that the contributions of the orbital and spin currents into the transverse $E1$ form factors at the momentum transfer ranging from $q = E_{excitation}$ up to $q \sim 0.5 - 0.7$ fm$^{-1}$ have different signs. The destructive interference in this $q$-region results in the minimum of the squared transverse $E1$ form factor [5]. The so-called "spin-flip" isovector $E1$ transitions $1l_{j=l+1/2} \rightarrow 1(l+1)_{j=l+1/2}$ are dominated by the spin-current contributions.

The comparison of the $q$-dependences for the single-particle transitions $1p_{3/2} \rightarrow 1d_{5/2}$ and $1p_{3/2} \rightarrow 1d_{3/2}$ is illustrated in the Fig.2. In the $q$-region where $1\hbar\omega$ transverse form factors for $1l_{j=l+1/2} \rightarrow 1(l+1)_{j=l+3/2}$ transitions are near their minima, the single particle $C1$ form factors reach their maxima. As a result, the observed at lower $q$ similarity between the shapes of $C1$ and $E1$ ($e,e'$) form factors should disappear when the momentum transfer $q$ grows up to 0.4 - 0.6 fm$^{-1}$.

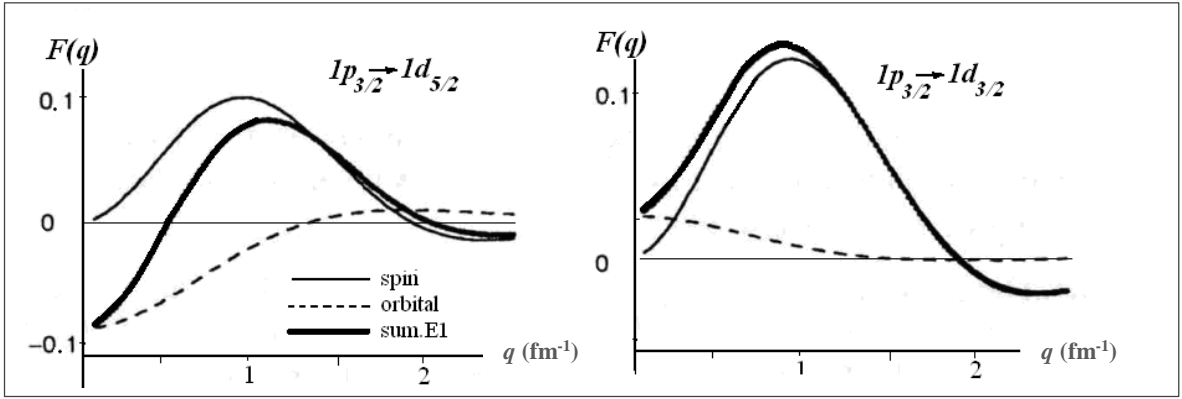

**Fig. 2.** Form factors for single particle $E1$ transitions $1p_{3/2} \rightarrow 1d_{5/2}$ and $1p_{3/2} \rightarrow 1d_{3/2}$ with separated spin and orbital modes.

Fig.3 compares the positions of $C1$ maxima on the $q$-axis with the positions of their respective $E1$ minima. The calculations were performed using HOWF. At higher momentum transfers the contribution of the spin operator grows and the role of the orbital operators decreases. As a consequence, the mean weighted energies of $E1$ resonances are shifted upwards with growing $q$. The mean energies of $C1$ response to the electroexcitation are almost stable at $q < 1.8$ fm$^{-1}$.

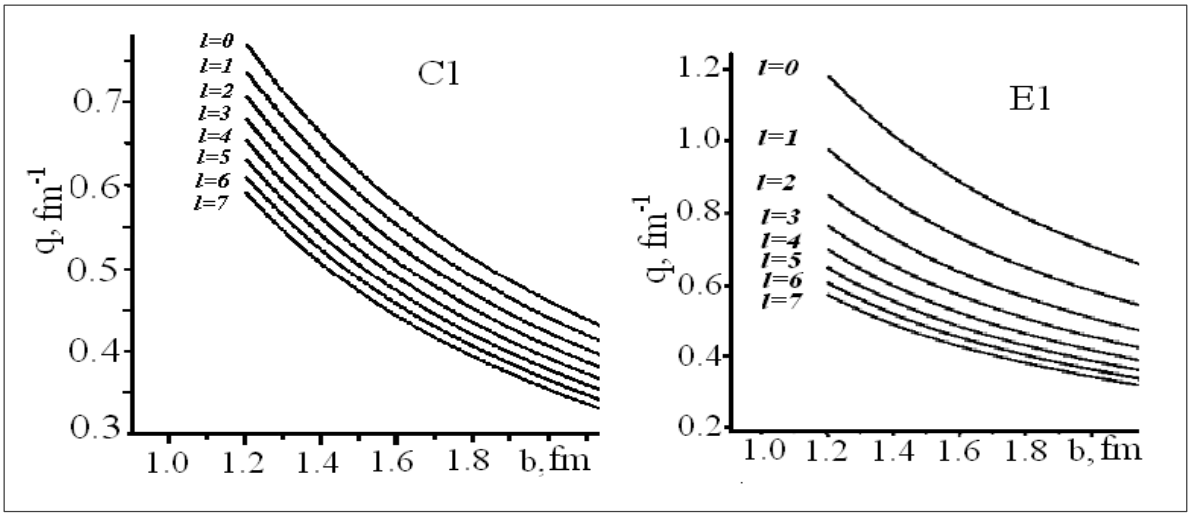

**Fig. 3.** Positions of maxima of $C1$ form factors (left ) and minima of $E1$ form factors (right) on the $q$ axis.

The effect of the destructive interplay between the spin and the orbital current components for the light nuclei $^{12}$C and $^{16}$O was observed at MAMI-A [6] when the $C1$ and $E1$ form factors of the electric dipole resonances were separated (see Fig.4). The disappearance of the $F_T$ form factor at $E \approx 21\text{-}22$ MeV is a result of the destructive interference between the spin and orbital currents at $q = 0.6$ fm$^{-1}$ in the $1p_{3/2} \to 1d_{5/2}$ transition shown in the Fig.2.

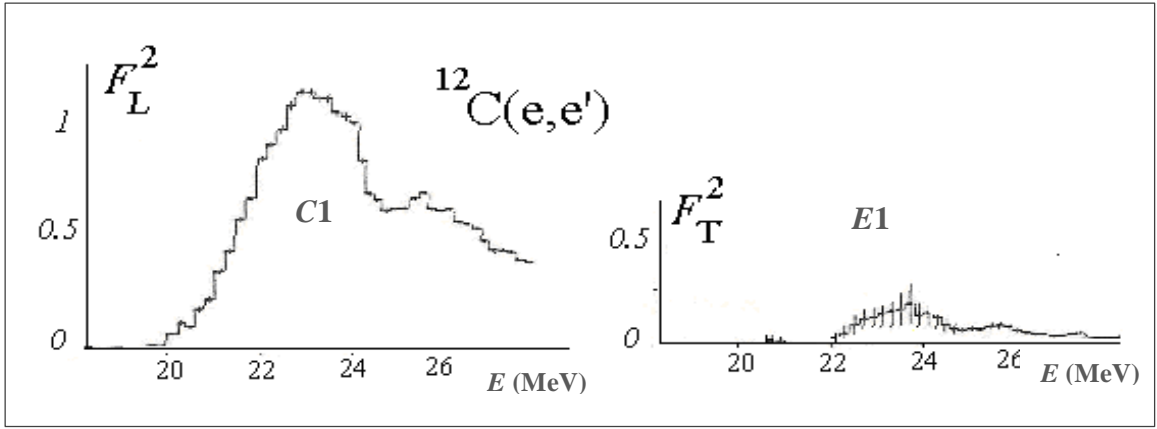

**Fig. 4**. $C1$ и $E1$ form factors for $^{12}$C nucleus at $q = 0.6$ fm$^{-1}$ [ 6 ].

The contrast behavior of $E1$ and $C1$ form factors for all $1l_{j=l+1/2} \to 1(l+1)_{j=l+3/2}$ transitions could be used as a tool for the identification of wave function configurations [7].

### 4. $M2$ : Magnetic quadrupole excitations

While the magnetic dipole ($M1$) modes of nuclear electroexcitation have been studied in detail for a number of stable nuclei, much less is known about the magnetic quadrupole ($M2$) resonances. $M2$ resonances have been investigated for a few light and medium nuclei at S-DALINAC [8-10 and references there ].

The excitation of the magnetic $M2$ resonances is mostly due to the spin-multipole operators containing both spin-dipole and spin-octupole parts corresponding to $[Y_1 \times \hat{\sigma}]^2$ and $[Y_3 \times \hat{\sigma}]^2$ terms, respectively. The first term dominates approximately at $0 < q < 1$ fm$^{-1}$ while the second term starts to play an important role at $q > 1$ fm$^{-1}$. Also the strength of the $M2$ mode has a significant contribution from the matrix element of the orbital operator $[Y_2 \times \hat{v}]^2$ as well. The orbital modes of $M2$ excitations were interpreted as the nuclear "twist" mode in [11].

The orbital mode completely vanishes in $1p_{3/2} \to 2s$ transitions of $M2$ excitations in $1p$-shell nuclei. The destructive interference between the spin and the orbital currents can be observed in $1p_{1/2} \to 1d_{5/2}$ and $1p_{1/2} \to 1d_{3/2}$ transitions. The transition $1p_{1/2} \to 1d_{3/2}$ is the most interesting one in terms of possible extraction of the "twist" mode because for this transition $\hat{O}_2^{mag} \sim g\left(\frac{1}{2} - y\right) - 1$. At $y = \left(\frac{bq}{2}\right)^2 \approx 0.5$ the spin mode disappears, which means that the $1p_{1/2} \to 1d_{3/2}$ transition is formed only by the "twist" mode. If the oscillator parameter $b$ ranges from 1.65 to 1.7 fm, this effect is achieved at the momentum transfer $q \sim 170$ MeV/c.

In $sd$-shell nuclei the $1\,\hbar\omega$ transitions from the $2s$ subshell do not contain any orbital-current contributions. The role of the "twist" mode for the $1d$-shell transitions is determined by the total angular momenta of the initial and final states of the valence nucleon. The orbital component plays the most essential role in the following three matrix elements: $\langle 1f_{7/2}\|\hat{O}_2^{mag}\|1d_{5/2}\rangle$ , $\langle 1f_{5/2}\|\hat{O}_2^{mag}\|1d_{5/2}\rangle$ and $\langle 1f_{5/2}\|\hat{O}_2^{mag}\|1d_{3/2}\rangle$, among which the latter appears to be the most promising for detecting the "twist" mode [12].

Fig.5 illustrates the contributions of the spin and orbital terms into the $M2$ form factors for two $1\,\hbar\omega$ transitions in $sd$-shell nuclei [13]. The renormalized value of the isovector constant $g$ used in these calculations is $g = 0.7g_{\text{free}}$ .

The destructive interference between the orbital and spin parts of the $M2$ operator reveals itself in the total disappearance of the $M2$ form factors at a certain value of momentum transfer $q$. Since the positions of these non-diffraction minima on $q$ axis depend on the configuration structure of the $M2$ peak, analyzing the $q$-dependence of $M2$ form factor could be helpful in identifying this structure.

The "twist" mode in the $M2$ electroexcitation is most likely to be detected when the matrix element of the sum of two spin operators (spin dipole and spin octupole) equals to zero. For the single-particle $1d_{3/2} \rightarrow 1f_{5/2}$ transitions the spin current contribution to $M2$ form factors disappears at $q \approx 0.8 - 1.0$ fm$^{-1}$. For $1d_{3/2} \rightarrow 1f_{7/2}$ transitions the orbital current dominates at a higher value of the momentum transfer: $q \approx 1.3$ fm$^{-1}$.

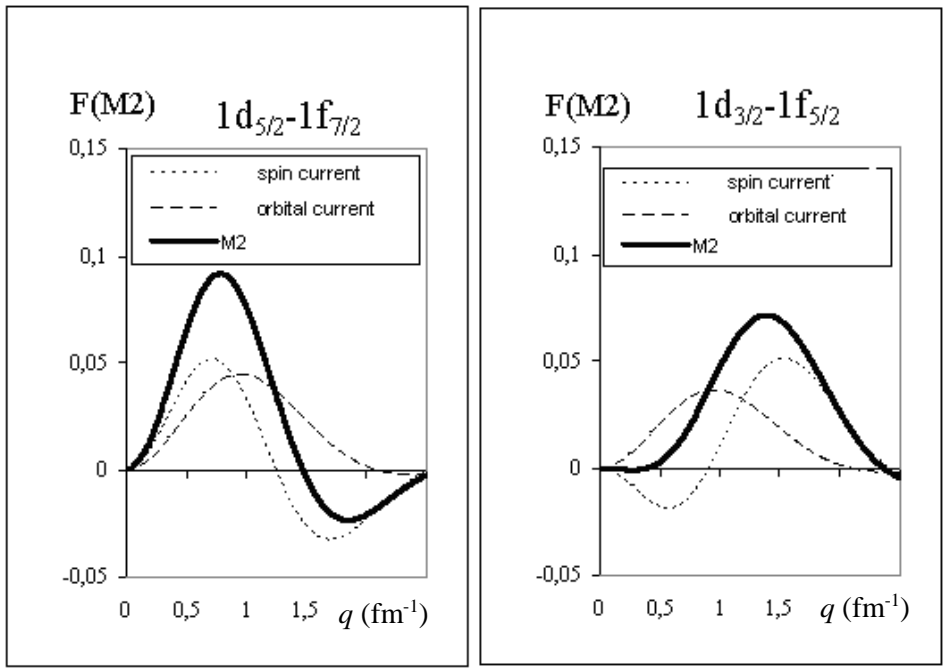

**Fig. 5.** Contributions of the spin and orbital currents into the form factors of $M2$ single-particle transitions.

The calculations of $M2$ states in $^{32}$S nucleus were performed within PCC SM [13 + references there] using the spectroscopic factors from [14]. The PCC basis for $M2$ states in $^{32}$S includes 38 configurations and takes into account all the states of $A$=31 nuclei with the spectroscopic factors $S > 0.1$. Out of those 38 configurations 36 correspond to the transitions from the $sd$-shell into the $fp$-shell and 2 configurations are built on the (1/2)$^-$ states of the residual nuclei and correspond to the transitions $1p_{1/2} \rightarrow 1d_{5/2}$ and $1p_{1/2} \rightarrow 1d_{3/2}$. Despite the small number of configurations built on these two transitions, they make a relatively significant contribution to the summed strengths of the $M2$ resonance because of a large spectroscopic factor $S$ $(1p_{1/2})$ for the pick-up reactions from $1p_{1/2}$ subshell: $S$ $(1p_{1/2})$ =1.6 [14], which accounts for 13% of the total $S$.

Fig.6 shows the results of these calculations at $q = 0.6$ fm$^{-1}$ compared with the experimental data obtained at S-DALINAC [10] at the excitation energies up to $E = 14$ MeV . Again, the renormalized value of the isovector $g$-factor used in the calculations is $0.7g_{\text{free}}$. The fragmentation of $M2$ strength that can clearly be observed in the Fig.6 is the result of spreading of the "hole" configuration among 12 states of $^{31}$S.

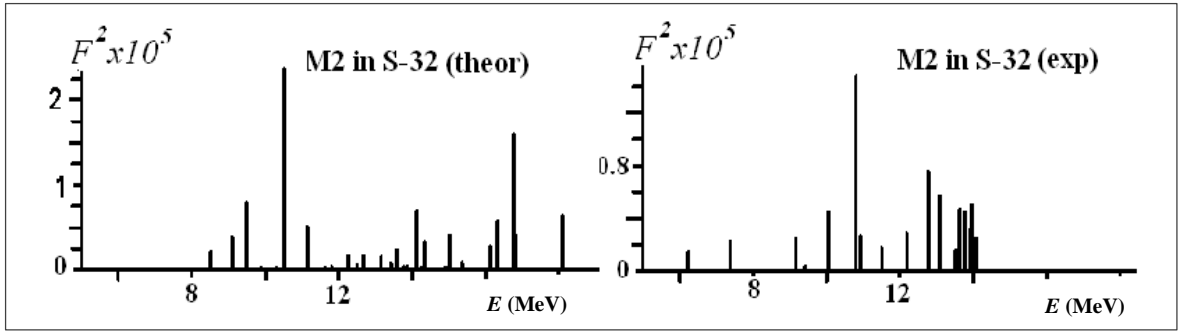

**Fig. 6.** Fragmentation of the M2 strength: comparison between theoretical and experimental results.

Fig.7 illustrates how the final result of the spin- and orbital-current contributions depends on the interplay between their signs. It can be observed that for the $M2$ states at $E < 10.6$ MeV the interference of the spin and orbital currents is mostly destructive. For the peak at $E \approx 10.6$ MeV the orbital part is larger than the spin part and their interference is constructive. As can be seen from the Fig.8, the $q$-dependence of the $M2$ form factor at $E \approx 10.6$ MeV calculated within the PCC SM model with $g = 0.7g_{\text{free}}$ is in a good agreement with the experimental data [10].

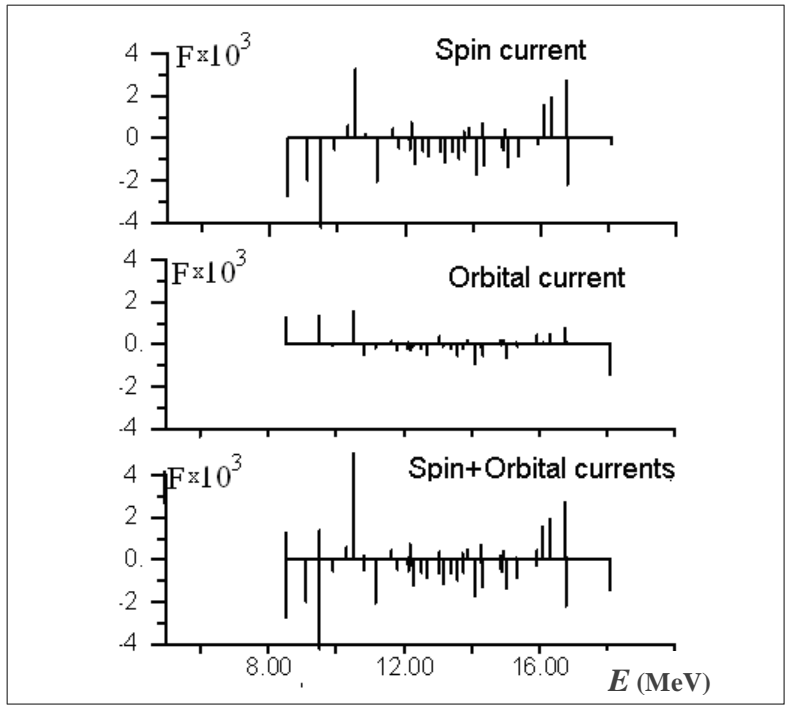

**Fig. 7.** Spin and orbital current contributions to the $M2$ form factor at $q = 0.6$ fm$^{-1}$ (the renormalization factor $g = 0.7g_{\text{free}}$ was used).

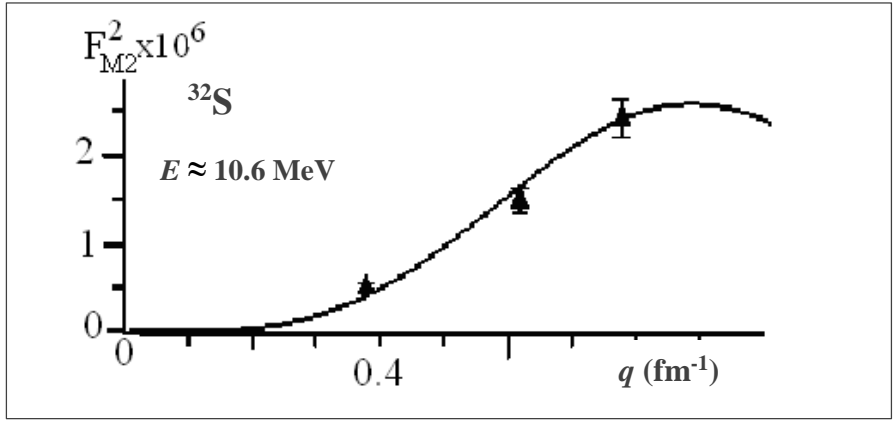

**Fig. 8.** $M2$ form factor for $^{32}$S at E ≈ 10.6 MeV. Solid line corresponds to the theoretical results calculated within PCC approach with $g=0.7g_{free}$ ; the experimental data are taken from [10].

## 5. Conclusions and outlook

The comparison of theoretical calculations for the spin and orbital currents contributing to the $(e,e')$ cross sections of $E1$ or $M2$ resonances is not the only method to separate the spin and orbital parts of nucleon currents. In [15] a possibility of comparative analysis of the experimental data obtained in the reactions of inelastic electron scattering $(e,e')$ and inelastic proton scattering $(p,p')$ is discussed. Individual resonance peaks belonging to purely orbital mode of the nuclear excitation can be identified from the comparative $(e,e')$ and $(p,p')$ experiments at close values of the momentum transferred to the nucleus. Such a separation of the orbital mode from the spin mode is possible due to the fact that the resonance peaks observed in the $(e,e')$ reactions are associated with both spin and orbital components of the nucleon current while in the $(p, p')$ scattering at small angles only the spin component plays an essential part.

The relative roles of the spin and orbital currents in the excitation of MGR are determined, for each particular nucleus, by the wave functions of its initial and final state. Therefore, the comparison of the theoretical results obtained within a certain model with the experimental data on the strength distribution of $E1$ and $M2$ resonances in $(e,e')$ and $(p,p')$ reactions is a very promising method to improve our understanding of the nature and structure of nuclear excited states.